\documentclass[mnsc,nonblindrev,copyedit]{informs2} % format for MS submission

\usepackage{natbib}
 \bibpunct[, ]{(}{)}{,}{a}{}{,}%
 \def\bibfont{\small}%
 \def\bibsep{\smallskipamount}%
 \def\bibhang{24pt}%
\TheoremsNumberedThrough     % Preferred (Theorem 1, Lemma 1, Theorem 2)
\ECRepeatTheorems

\EquationsNumberedThrough    % Default: (1), (2), ...
\MANUSCRIPTNO{{ \today}}

\begin{document}
%%%%%%%%%%%%%%%%

%\setcounter{page}{1}
%\DOI{10.1287/xxxx.0000.0000}%

\RUNAUTHOR{Huang, Wan and Chen}

\RUNTITLE{An Active Margin System for Margin Lending Transactions}

% \TITLE{Bundling Information Goods of Decreasing Value}
\TITLE{An Active Margin System and its Application in Chinese Margin Lending Market}

\ARTICLEAUTHORS{%
\AUTHOR{Guanghui Huang}
\AFF{College of Mathematics and Statistics, Chongqing University,
  Chongqing 401331, China, \EMAIL{hgh@cqu.edu.cn}}
   %, \URL{}}
\AUTHOR{Jianping Wan}
\AFF{College of Mathematics and Statistics, Huazhong University of Science and Technology, Wuhan 430074, China, \EMAIL{hust\_jp\_w@yahoo.com.cn}}
\AUTHOR{Cheng Chen}
\AFF{College of Mathematics and Statistics, Chongqing University,
  Chongqing 401331, China
  %\EMAIL{m.arinella@adult.ufp.edu}
  }
} % end of the block

\ABSTRACT{%
In order to protect brokers from customer defaults in a volatile market, an active margin system is proposed for the transactions of margin lending in China. The probability of negative return under the condition that collaterals are liquidated in a falling market is used to measure the risk associated with margin loans, and a recursive algorithm is proposed to calculate this probability under a Markov chain model. The optimal maintenance margin ratio can be given under the constraint of the proposed risk measurement for a specified amount of initial margin.  An example of such a margin system is constructed and applied to $26,800$ margin loans of 134 stocks traded on the Shanghai Stock Exchange. The empirical results indicate that the proposed method is an operational method for brokers to set margin system
with a clearly specified target of risk control.

% Enter your abstract
}%

%\KEYWORDS{deterministic inventory theory; infinite linear programming duality;
%  existence of optimal policies; semi-Markov decision process; cyclic schedule}

\KEYWORDS{margin lending; initial margin; maintenance margin;
           mandatory liquidation; Markov chain}
%\HISTORY{This paper was
%first submitted on April 12, 1922 and has been with the authors for
%83 years for 65 revisions.}

\maketitle
%%%%%%%%%%%%%%%%%%%%%%%%%%%%%%%%%%%%%%%%%%%%%%%%%%%%%%%%%%%%%%%%%%%%%%

% Samples of sectioning (and labeling) in MNSC
% NOTE: (1) \section and \subsection do NOT end with a period
%       (2) \subsubsection and lower need end punctuation
%       (3) capitalization is as shown (title style).
%
%\section{Introduction.}\label{intro} %%1.
%\subsection{Duality and the Classical EOQ Problem.}\label{class-EOQ} %% 1.1.
%\subsection{Outline.}\label{outline1} %% 1.2.
%\subsubsection{Cyclic Schedules for the General Deterministic SMDP.}
%  \label{cyclic-schedules} %% 1.2.1
%\section{Problem Description.}\label{problemdescription} %% 2.

% Text of your paper here

\section{Introduction} % the scale of Japan, USA, China and Australia
Using securities as collateral to get credit from a broker is called
buying on margin or margin lending. This transaction
 gives
investors the
possibility to buy more stocks than normal
and
in that way
increases profits or losses \citep{Peter2000,Peter2001a,Peter2001b, ricke_2003}.
Margin lending had been prohibited in China between 1996 and 2009, because of the absence of regulation and high default rates on both of the Shanghai Stock Exchange (SSE) and Shenzhen Stock Exchange (SZSE). However,  margin lending  was  officially allowed at both of the exchanges  on
March 31, 2010. The development of Chinese margin lending markets motives this research for risk measurement and margin setting for margin lending transactions.

% margin lending market should be regulated
Although margin lending provides the possibility to make  financial
markets more efficiently, it has been regarded as a source of instability of stock markets for a long time. In the 1920s, the brokers in the United States would lend as much as 90 percent of the money that customers needed to purchase common stocks, leaving only 10 percent equity margin to cushion declines in stock prices.
When the stock market started to plummet, many individuals received margin calls. They had to deliver more money to their brokers or their shares would be sold. Since many individuals did not have the equity to cover their margin positions, their shares were sold, causing further market declines and further margin calls. \citet{galbraith1954}, \citet{hsieh1990} indicated that low margin requirement was  one of the major contributing factors which led to the Stock Market Crash of 1929, which in turn contributed to the Great Depression.
Therefore the reduction in the frequency of margin calls under an acceptable level of risk may be beneficial to the stability of stock market.

In order to restrain the level of speculation and  protect investors and brokers,
the United States Congress gave the Federal Reserve System the power to control margin requirements in 1934. The initial margin requirement was changed 22 times between 1934 and 1974, and it has been fixed at $50\%$ since 1974.
Shiller claimed in a {\it Wall Street Journal} article that the stock market crash of 1987 and the stock market boom in the late 1990s led to calls for the return to an active margin policy which sets a minimum equity
position on the date of a credit-financed security transaction \citep{shiller2000, ricke_2003}. On the other hand, Bartlett argued that although the adjustment of margin policy can not impact the volatility of stock market, brokers should set their own house margins to react to the changes of stock market, which can protect themselves from customer defaults on margin loans \citep{shiller2000,Peter2001b}. Therefore the purpose of this paper is not to manage the volatility of stock prices, nor to restrain the level of speculations in stock market, but to construct a model that can be used by a broker to set margin requirements with a clearly specified target.

The margin system of Chinese margin lending markets is composed of initial margin requirement, maintenance margin requirement and mandatory liquidation, which is the line of defence for the brokers against the risk associated with the transactions of margin loans. \citet{Argiriou} pointed out that although supervising authorities give recommendations or limits for margin requirements, there is no
information on how these recommendations are derived. In order to construct a model that can be used by a broker to set margin requirements with a clearly specified target,
an alternative risk measurement is proposed in this paper.
It is supposed that investors will default whenever they receive their first margin calls.
Pursuant to the regulations of the SSE and SZSE,
brokers must sell the collaterals within two trading days,
when investors do not meet their margin calls. Considering the uncertainty in  market value of collateral to be liquidated,
it is not known whether the margin lending transaction can make a positive profit until mandatory liquidation takes place.
From this point, the risk of a margin lending transaction can be described by the
probability of negative return under the condition that the first margin call has been issued, which is called the conditional probability of negative return (CPNR) in the following sections. The stop-loss function of mandatory liquidation is taken into account to measure the risk of margin lending transactions.

% the determination of initial margin must be wrote
The value of CPNR is determined by two types of factors. One is the margin system, including the amount of initial margin and the required maintenance margin ratio. The second is  the market situation, including
 the current stock price, riskless interest rate, rate of margin loan, and the historical evolution of stock prices. The value of CPNR may vary frequently with  the changes of stock prices  under an invariant margin system, whose requirements of initial and maintenance margin ratios are unchanged  to different market situations. In order to control the risk faced by brokers  below an acceptable  level, the optimal initial and maintenance margin requirements should be active with respect to the changes of stock market.  CPNR provides a possibility to manage the risk exposure of brokers through a well-designed margin system.

 For a particular market situation, those
 margin systems with the same value of CPNR are called indifference margin systems. The optimal margin system with respect to some specified purposes, such as minimizing opportunity cost, minimizing the rate of default, minimizing the frequency of margin calls and other purposes involved in practice, can be found in the indifference set deduced from the constraint of CPNR.  Using CPNR as the risk measure, the resulted margin system can be explained more comprehensibly than those margin systems just following the recommendations of authorities.

% math needed
% math model of underlying asset price
In order to calculate the value of  CPNR, a mathematical model of stock price is needed. The geometric Brownian motion is used to construct  margin system
for futures market \citep{Figlewski}, however, whose normality assumption is
found to be inappropriate for stock prices \citep{warshawsky1989}.
The extreme value theory (EVT) is another favorable model to set margin system of futures market, including \citet{Booth}, \citet{Cotter}, \citet{Broussard} and the references therein. On the other hand, stock returns are found to be autocorrelated by  \citet{Famma}, \citet{Potreba}, \citet{jegadeesh1991} and  other authors,
which can not be described by EVT due to its basic assumption that stock returns are independent and identically distributed (i.i.d.). Another popular working model is GARCH model, such as \citet{Lam} and the references therein. However, \citet{Cotter} documented that the fat tail of returns is usually ignored by GARCH model. In order to describe the movements of stock price process without those distributional assumptions, a simple Markov chain model is adopted in this paper,  and a recursive algorithm for CPNR is proposed
in the following sections.

% the cost of invest must be involved
 In order to investigate the performance of the margin system deduced by CPNR, an example of such a margin system is constructed  in this paper, which is chosen from the indifference set with CPNR=0.05 using a least squares method. The resulted margin system is applied to 134 stocks listed in the SSE 180 Index, where
 200 margin loans with 30-day period are constructed for each stock to test the prudentiality of margin system. A stock is said to pass the out-of-sample test, when its  frequency of negative returns among those 200 margin loans is less than or equal to 0.05. There are 119 stocks which pass the out-of-sample test in our empirical investigations
\footnote{The prices used in the empirical investigations are downloaded from the website of SSE, and the last date of the sample is April 30, 2010.
   There are more 8 stocks which can pass the out-of-sample test, when the sample size used to estimate the transition matrix is increased from 800 to 1,500. Among the remaining 7 stocks, there are still 4 stocks which can not pass the test even under the margin system recommended by the stock exchanges, due to the abnormal price movements.}.

A more detailed analysis is applied to those 119 stocks which have passed the out-of-sample test. Under the proposed margin system, the average initial margin ratio of $23,800=119 \times 200$ margin loans considered in this paper is around $57\%$, and the average maintenance margin ratio is around $121\%$. The number of margin calls
for each stock among 200 margin loans is averagely $22.30$ under the required margin system, which is recommended by the China Securities Regulatory Commission (CSRC) and the two exchanges,  and its counterpart under the proposed margin system
is averagely $0.97$. The cost of each margin loan, which is the terminal time value of the initial margin and the additional capital to meet all of the margin calls during the period of the loan, is averagely $9.97$ yuan under the required margin system, and its counterpart under the proposed margin system is $11.08$ yuan. It can be concluded that the frequency of margin calls is reduced by $95.7\%$ with more $11.1\%$ cost under the proposed margin system. Those empirical investigations indicate that the idea of CPNR can be used to measure the risk associated with the transaction of margin loan, and the margin system deduced from the constraint of CPNR is valuable for brokers to set their own house margins.

The rest of this paper  is organized as follows.
The risk measurement for
Chinese margin lending market is discussed in section 2, and the
algorithm to realize this risk measurement is given in section 3. A relatively simple margin system is proposed under the constraint of CPNR in section 4. The performance of the proposed margin system is investigated in section 5. The discussions and conclusions are given in section 6.

\section{Risk Measurement}

\subsection{Margin System in Chinese Market}

Pursuant to the
regulations of the SSE and SZSE,
an amount of initial margin must be deposited to broker
on the date of margin lending transaction.  The
ratio of the initial margin to the market value of the stocks
purchased by the loan is called {\it{initial margin ratio}}, which should be larger than or equal to $50\%$, the minimum initial
margin ratio determined by the two stock exchanges and CSRC.

Initial margin and
all of the stocks purchased by the loan are kept in the account of collaterals.
The ratio of the market value of collaterals to the market
value of the loan is called {\it {maintenance margin ratio}}, which
should be larger than or equal to $130\%$, the minimum maintenance margin ratio
determined by CSRC and the two stock exchanges.

Maintenance margin ratio is marked to market by brokerage house,
and broker must issue a margin call when the current value of maintenance margin ratio is below the required minimum value. Investors now either have to increase the margins that they have deposited or close out their positions. If they do none of
these, brokers can sell their collaterals to meet the margin
calls, which is called {\it mandatory liquidation} in Chinese markets. Mandatory
liquidation is the last line of defence against the potential losses of margin lending transactions. Pursuant to the regulations of the SSE and SZSE, the collaterals should be liquidated within two trading days if investors do not meet their margin calls.

\subsection{Conditional Probability of Negative Return}
In order to explain  in detail the idea of risk measurement for margin lending transaction,
a simple margin loan is constructed in this paper, where
the initial margin deposited to  broker is only cash without other securities, and there is
only one share of stock purchased by the loan. Let $Q_0$ denote the
amount of initial margin, and $P_i$ denote the market price of the
purchased stock on the ith
trading day $(i=0,1,2,\cdots,T)$,  where $T$ is the expiration date of the loan.
The market value of the initial margin can be divided into two parts on the ith trading day, i.e.
\begin{equation}
Q_{0}\left( 1+r \right)^{i}= \Sigma_{i}+L_{i},
\end{equation}
where $r$ is the risk free one-day interest rate, $\Sigma_{i}$ is the
required amount of initial margin, and $L_{i}$ is the remaining  margin after fulfilling the requirement of initial margin. The market value of the loan on the ith day is
\begin{equation}
P_{0}\left( 1+R \right)^{i},
\end{equation}
where $R$ is the one-day loan interest rate.

Let $m_{0}$ denote the initial margin ratio on the date of margin lending transaction, i.e.
\begin{equation}
m_{0} = \frac{Q_0}{P_0},
\end{equation}
which should satisfy the requirement of maintenance margin, i.e.
\begin{equation}
\frac{Q_0+P_0}{P_0}=\frac{m_0 P_0 + P_0}{P_0} \ge w,
\end{equation}
where $P_0$ is the market value of the loan on the date of transaction, which is the same as the  stock price on the same date.
And $w$ is the required minimum maintenance margin ratio. The following proposition
can be used to determine whether the initial margin is adequate on the date of  transaction.
\begin{proposition}\label{initialmargin}
Let $m_0$ be the initial margin ratio on the date of transaction, and $w$ denote the required minimum maintenance margin ratio, then the initial margin is adequate iff
\begin{equation}
m_0 + 1 \ge w.
\end{equation}
Otherwise, investor should deposit more margin to  broker to fulfill the requirement of maintenance margin.
\end{proposition}

\begin{remark}
\cite{Peter2000}  showed that the aggregate amount of debit balances at broker-dealers per dollar of potential margin debt never exceeded $50\%$ between 1985 and 2000 in the U.S. market. This observation indicates that the amount of initial margin deposited by investors on the date of transactions are much more than the required minimum amount determined by the authorities. Therefore we focus on the amount of initial margin to measure the risk associated with a margin lending transaction, rather than the required minimum initial margin ratio in the following sections.
\end{remark}

When the requirement of maintenance margin is fulfilled on the date of  transaction, the amounts of the required margin and the remaining margin satisfy the following proposition before the first margin call.
\begin{proposition} \label{marginwithoutm}
Let $w$ denote the required minimum maintenance margin ratio, $P_i$ denote the closing stock price on the ith day,  $R$ denote the one-day loan interest rate, $r$ denote the one-day riskless interest rate, and
$Q_0$ be the amount of initial margin, then the amounts of the required margin and the remaining margin satisfy
\begin{eqnarray}
\Sigma_i & = & w P_0 \left(  1+R \right)^{i} - P_i,\\
L_i      & = & Q_0 \left( 1+ r \right)^{i}- w P_{0} \left( 1+ R \right)^{i} + P_i.
\end{eqnarray}
\end{proposition}

\begin{remark}
The time of margin call is determined by the amount of remaining margin. From Proposition $\ref{marginwithoutm}$, we can find that the amount of remaining margin is determined by several factors, including the amount of initial margin $Q_0$, minimum maintenance margin ratio $w$, market value of the loan on the date of transaction $P_0$, two interest rates $r$ and $R$, and the evolution of stock  prices during the period of the loan. The required minimum initial margin ratio is not one of  the key factors to determine the time of margin call.
\end{remark}

When there is no more remaining margin,  broker must issue a margin call, i.e.
\begin{equation}
L_{\tau} = Q_{0} \left( 1+ r \right)^{\tau} - w P_{0} \left( 1+R \right)^{\tau}+ P_{\tau} \le 0,
\end{equation}
where $\tau$ is the stopping time of the first margin call during the period of the loan, i.e.
\begin{equation}
\tau = \min \left\{ i \in \left\{1,2,\cdots,T\right\}:
  Q_0 \left( 1+ r \right)^{i} - w P_{0} \left( 1+R \right)^{i}+ P_{i} \le 0
    \right\}.
\end{equation}
The condition of first margin call can be rewritten as
\begin{equation}
\mathrm{C} =
 \left\{
 P_{\tau} < wP_{0} \left( 1+R \right)^{\tau} - Q_0 \left( 1+r \right)^{\tau}
 \right\}.
\end{equation}

Pursuant to the regulations of the SSE and SZSE, collaterals should be liquidated within two trading days, if investors do not meet their margin calls. For simplicity, it is supposed that investors would default whenever they receive their first margin calls, and the mandatory liquidations take place within one trading day in the following sections, i.e.
\begin{equation}
\tau^* = \min \left\{ \tau + 1, T \right\},
\end{equation}
where $\tau^*$ is the date of mandatory liquidation. After liquidation, the return of a margin lending transaction is
\begin{equation}
P_{\tau^*} + Q_0 \left( 1+r \right)^{\tau^*} - P_0 \left(  1+ R \right)^{\tau^*},
\end{equation}
which is negative  when the following condition is satisfied
\begin{equation}
\mathrm{N} =  \left\{
P_{\tau^*} < P_0 \left(  1+ R \right)^{\tau^*} - Q_0 \left( 1+r \right)^{\tau^*}
\right\}.
\end{equation}
The conditional probability of negative return (CPNR) can be denoted as
\begin{equation}
\mathrm{CPNR} = \mathrm{Prob}
\left\{
N |
 C
\right\},
\end{equation}
where $\mathrm{Prob}\left\{ N | C \right\}$ denotes the probability of N under the condition that C has happened.
If the interest rate of the loan is
 set to be the same as the riskless interest rate for simplicity, the formula of CPNR can be rewritten as
 \begin{equation}\label{CPNR}
\mathrm{CPNR}=\mathrm{Prob}\left\{
P_{\tau^*}<\left(P_0-Q_0\right)\left(1+r\right)^{\tau^*}|
P_{\tau}<\left(wP_0-Q_0\right)\left(1+r\right)^{\tau}
\right\},
\end{equation}
which is the definition of CPNR to be used in the following sections.

\subsection{CPNR and Risk Control}
% active margin  % cpnr and risk
From (\ref{CPNR}) we can find that CPNR is a function of several factors, including the amount of initial margin $Q_0$, current stock price $P_0$,  minimum maintenance margin ratio $w$, riskless interest rate $r$, and the historical dynamics of the stock prices. Suppose the margin system is an invariant system, whose requirements of  initial and maintenance margin ratios are unchanged to different market situations.
Let $m$ denote the invariant initial margin ratio, and the amount of initial margin deposited by the investor is $Q_0=m P_0$, then the conditional probability of negative return is
\begin{equation}\label{invcpnr}
\widehat{\mathrm{CPNR}}=\mathrm{Prob}\left\{
P_{\tau^*}<\left(1- m\right) P_0 \left(1+r\right)^{\tau^*}|
P_{\tau}<\left(w- m\right) P_0 \left(1+r\right)^{\tau}
\right\}.
\end{equation}
From (\ref{invcpnr}) we can find that the value of $\widehat{\mathrm{CPNR}}$ is determined by the current stock price $P_0$ and the historical evolution of stock prices under this invariant margin system, which means that the risk faced by brokers in their margin lending transactions is dynamic with respect to the changes of stock prices. In order to control the risk to an acceptable level, an active margin system is needed in the practice of margin lending transactions.

The risk of margin lending transaction can be explained more intuitively with the help of CPNR.
For example, CPNR equals to 0.05, which means that if the broker issues 100 margin calls for the margin loans under the same market situation, there are at most 5 times that he/she yields a negative return after mandatory liquidation, due to the protection of  margin system. It is more possible to have a negative return  when the value of CPNR becomes larger. The value of CPNR can be controlled through the adjustment of $Q_0$ and $w$, which
gives the possibility to manage the risk associated with margin lending transactions by a well designed margin system.

As an alternative risk measurement, CPNR
does not only take into account the potential losses of investors during the period of loans, but also emphasize on the market value of collaterals and the stop-loss function of mandatory liquidation, which is different from those risk measurements for futures market, where the margin systems are designed to cover the potential losses in the next trading day.

\section{CPNR under Markov Chain}
\subsection{Construction of Markov Chain}
The value of CPNR will be calculated under a Markov chain model in the following sections.
A Markov chain is a random process that jumps from one state to another, whose next state depends only on the present state. In order to construct a Markov chain model, the space of states must be identified, and the probabilities of transitions must be estimated from the historical observations of the chain. The daily closing prices are used to finish these two tasks in the following sections.

A relatively simple method is used to reconstruct the Markov version of stock price process in this paper.  The observed price data are sorted in order of increasing, and every g different prices are regarded as one state, where g is called the number of a group in this paper.  And the sample size used to construct the Markov chain model is called the depth of memory. In order to estimate the transition probabilities more accurately, a larger sample size is needed for a fixed g. However, when more historical data are used to reconstruct the evolution of stock prices,
the events which had happened a  long time ago will disturbance the prediction of current tendency. There should be a trade off between the estimation of transition probabilities and the prediction of future tendency.
The depth of memory to be used in the following sections is 800.

The estimation of transition probability is more straightforward when the state space is fixed. Suppose there are n elements in the state space, denoted as $\{s_1,s_2,$ $\cdots,$ $s_n\}$, the probability of transition from the ith  to the jth state
after one step is estimated by
\begin{equation}
\hat{p}_{ij}(1)= \frac{f_{ij}}{f_{i\cdot}},
\end{equation}
where $f_{ij}$ is the observed number of transitions from the ith to the jth state  after one step, and $f_{i\cdot}=\sum_{j=1}^{n} f_{ij}$ is the total number of stock prices falling into the ith state.  The one step transition matrix can be estimated by
\begin{displaymath}
\mathbf{P(1)} = \left( \begin{array}{ccc}
\hat{p}_{11}(1) & \hat{p}_{12}(1) & \ldots \quad  \hat{p}_{1n}(1)\\
\hat{p}_{21}(1) & \hat{p}_{22}(1) & \ldots \quad  \hat{p}_{2n}(1)\\
\vdots & \vdots & \vdots\\
\hat{p}_{n1}(1) & \hat{p}_{n2}(1) & \ldots \quad  \hat{p}_{nn}(1)\\
\end{array} \right).
\end{displaymath}
Denote $\mathbf{P(n)}$ the n-step transition matrix, which can be estimated by \begin{equation}
\mathbf{P(n)}=\mathbf{P^n(1)},
\end{equation}
where the chain has been supposed to be stationary.

\begin{figure}
\caption{An Example of Transition Matrix} \label{zhuanyijuzhen}
\begin{center}
\includegraphics[width=0.9\textwidth]{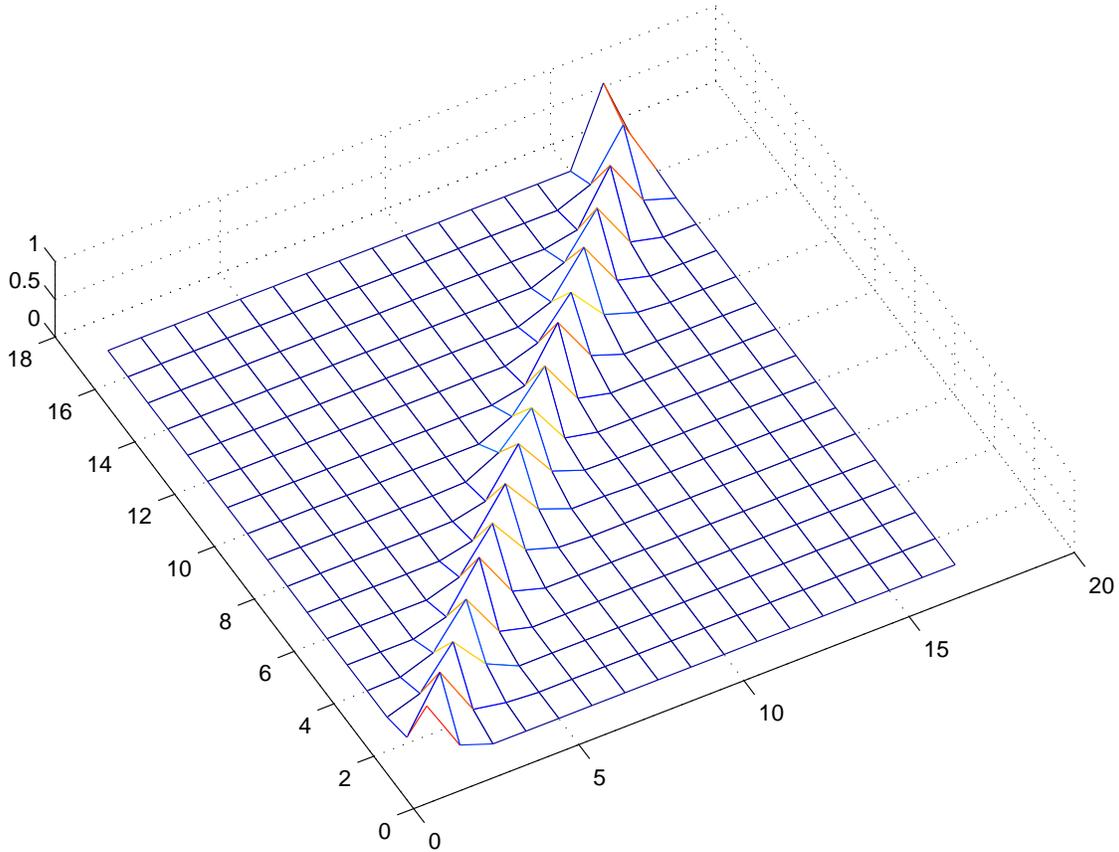}
\end{center}
\end{figure}

An example of transition matrix is given in Figure \ref{zhuanyijuzhen}, where the closing prices of SH600018 between
July 18, 2007 and November 12, 2010 are used to estimate the one step transition matrix.  From this figure we can find that all of the entries of the transition matrix are concentrated around the diagonal line, which means that the state of the next price is always closed to the state of the previous one. This observation can be regarded as the evidence of Markov property of stock prices.

\subsection{Test for Markov Property}
\citet{Billingsley} documented several statistics for Markov chain, where a
test for the null hypothesis that the process $\left\{ X_1,X_2,\cdots,X_{N+1} \right\}$  is an i.i.d. sequence within the hypothesis that it is a first-order Markov chain was discussed in detail \footnote{
\citet{Billingsley} documented that if H is a hypothesis contained in the larger hypothesis $H'$, we will speak of testing H within $H'$, rather than of testing H against alternatives in $H'-H$.
}.
 Suppose there are $n$ elements in the state space, a statistic $\chi^2$ is given by
\begin{equation}
\chi^2=\sum_{i,j}\frac{\left(f_{ij}-f_{i.}f_{.j}/N\right)^2}{f_{i.}f_{.j}/N},
\end{equation}
whose asymptotical distribution is a chi-square distribution with $(n-1)^2$  degrees of freedom, where $f_{.j}=\sum_{i=1}^n f_{ij}$.

\begin{table}
\vskip 0.3cm
\caption{Test for Markov Property with Five Stocks Traded on the SSE}\label{tablemarkov}
\begin{center}
\begin{tabular*}{15cm}{@{\extracolsep{\fill}}rccc}
\hline
Stock Code	&	Degrees of Freedom  	&	Observed $\chi^2$ Value	    & P 	 Value     \\
\hline
sh600000	&	676	&	20811.63644	&	0.000	\\
sh600005	&	441	&	18301.43754	&	0.000	\\
sh600007	&	484	&	18863.01088	&	0.000	\\
sh600011	&	256	&	15058.27575	&	0.000	\\
sh600018	&	256	&	15472.79408	&	0.000	\\
\hline
\end{tabular*}
\end{center}
\vskip 0.5cm
{Notes. The closing prices from
July 18, 2007 to November 12, 2010 are used to test the Markov property. For each stock reported in this table,  the null hypothesis that the sequence of stock prices is an i.i.d. sequence is rejected, and the hypothesis that the sequence is a Markov chain is accepted.}
\end{table}

An example of test for Markov property is given in Table \ref{tablemarkov}, where five stocks are randomly chosen from the SSE 180 Index.  The closing stock prices between July 18, 2007 and November 12, 2010  are used to estimate the transition count matrix $F=(f_{ij})$. From Table \ref{tablemarkov}, it can be found  that the null hypothesis that the sequence of stock prices is an i.i.d. sequence is rejected, and the hypothesis that the sequence is a Markov chain is accepted. This observation indicates that the Markov chain model constructed in this paper can be used as an alternative model for stock price processes.

\subsection{Stopping Time and CPNR}
The event that the broker issues a margin call for the first time during the period of the loan is denoted as
\begin{equation}
B= \left\{  P_{\tau} < \left( w P_0 - Q_0 \right) \left(  1+r \right)^{\tau}  \right\},
\end{equation}
and the margin lending transaction brings the broker a negative return after mandatory liquidation is denoted as
\begin{equation}
A = \left\{  P_{\tau^*} < \left(  P_0 - Q_0 \right) \left( 1+r \right)^{\tau^*}  \right\}.
\end{equation}
The conditional probability of negative return (\ref{CPNR}) is rewritten as
\begin{equation}
\mathrm{CPNR} = \mathrm{Prob} \left\{ A | B  \right\} = \frac{\mathrm{Prob}\left\{AB\right\}}{\mathrm{Prob}\left\{B\right\}},
\end{equation}
 provided that the denominator is non zero. On the other hand, when the denominator is zero, the value of CPNR is also zero following its  definition.
In order to calculate the value of CPNR,  the numerator and denominator should be calculated
respectively.

Let
\begin{equation}
D_i= \left\{  P_{i} < \left( w P_0 - Q_0 \right) \left(  1+r \right)^{i}  \right\},
\end{equation}
where $i=1,2,\cdots,T$. And let $B_i$ denote the event that the broker
issues the first margin call on the ith day  during the loan period, such that $B_i \bigcap B_j = \emptyset $
if $i  \neq j$. The probability of the broker to issue a margin call
for the first time during  the period of the loan can be
calculated by the following proposition.
% the proposition of P{B}
\begin{proposition}[A recursive algorithm for $\mathrm{Prob} \left\{ B \right\}$]
Suppose the current stock price $P_0$  belong to the state $s_h$, where $h \in \left\{1,2,\cdots,n\right\}$,  then the probability that the broker issues a margin call for the first time during the period of the loan is given by
\begin{equation}
\mathrm{Prob}(B)=\sum_{t=1}^T \mathrm{Prob}(B_t),
\end{equation}
where
\begin{equation}
\mathrm{Prob}(B_t)=\mathrm{Prob}(\overline{D}_1)
\mathrm{Prob}(\overline{D}_2|\overline{D}_1)\cdots{\mathrm{Prob}(\overline{D}_{t-1}|
\overline{D}_{t-2})}\mathrm{Prob}({D_{t}}|\overline{D}_{t-1}),
\end{equation}
and
\begin{equation}
\left\{ \begin{aligned}
         \mathrm{Prob}(\overline{D}_1)&=1-\sum_{i=1}^{k_1}\hat{p}_{hi}(1),\\
         \mathrm{Prob}(D_m|\overline{D}_{m-1})&
         =\frac{\sum_{i=k_{m-1}+1}^n
                  \sum_{j=1}^{k_m}\hat{p}_{hi}(m-1)
                   \hat{p}_{ij}(1)}{\sum_{i=k_{m-1}+1}^{n}
                    \hat{p}_{hi}(m-1)},\\
  \mathrm{Prob}(\overline{D}_m|\overline{D}_{m-1}) &
         = 1-\mathrm{Prob}(D_m|\overline{D}_{m-1}),\\
         \end{aligned} \right.
                          \end{equation}
$m=2,\cdots,t$, and $k_m$ is the largest state index which satisfies
\begin{equation}
 k_m = \max \left\{ k \in \left\{ 1,2,\cdots,n \right\} : q_{k} < (wP_0-Q_0)(1+r)^{m} \right\},
\end{equation}
where $q_k$ is the representative price level for the kth state.
\end{proposition}

\begin{remark}
$\mathrm{Prob}\left\{ B_t \right\}$ can be calculated in a recursive manner.
Suppose the previous conditional
 probabilities have been calculated, i.e.
\begin{equation}
\mathrm{Prob}(\overline{D}_1)
\mathrm{Prob}(\overline{D}_2|\overline{D}_1)\cdots{\mathrm{Prob}(\overline{D}_{t-1}|
\overline{D}_{t-2})}
\end{equation}
is known, then the only new probability needed for $\mathrm{Prob}\left\{ B_t \right\}$ is
\begin{equation}\label{newprob}
\mathrm{Prob}({D_{t}}|\overline{D}_{t-1}),
\end{equation}
which is given by Proposition 3. The remaining probability of (\ref{newprob}) is
\begin{equation}
\mathrm{Prob}(\overline{D}_{t}|\overline{D}_{t-1})
=
1-\mathrm{Prob}({D}_{t}|\overline{D}_{t-1}),
\end{equation}
which can be used to  calculate $\mathrm{Prob}\left\{ B_{t+1} \right\}$
 in the next step.

If  $\mathrm{Prob}\left\{ \overline{D}_{t-1} \right\}=0$,
the price process has entered  a state where the broker should issue a margin call with probability one, such that there is no possibility for the broker to issue the first  margin call in the following days, therefore the probabilities $\mathrm{Prob}\left\{ B_j \right\}$ must be zeros for $j=t,t+1,\cdots,T$.
\end{remark}

\begin{proposition}[A recursive algorithm for $\mathrm{Prob}(AB)$]
Suppose the current stock price $P_0$ belong to state $s_h$, then the probability that the two events A and B have happened simultaneously is given by
\begin{equation}
\mathrm{Prob}(AB)=\sum_{t=1}^T \mathrm{Prob}(AB_t),
\end{equation}
where
\begin{equation}
\begin{aligned}
         \mathrm{Prob}(AB_t)= &\mathrm{Prob}
         (\overline{D}_1)\mathrm{Prob}(\overline{D}_2|\overline{D}_1)
         \cdots \\
         &
         {\mathrm{Prob}(\overline{D}_{t-1}|
          \overline{D}_{t-2})}\mathrm{Prob}({D_{t}}|\overline{D}_{t-1})
          \mathrm{Prob}(A|D_t),
          \end{aligned}
\end{equation}
and
\begin{eqnarray}
\mathrm{Prob}(A|D_t)=
\left\{
 \begin{aligned}
           & \frac{\sum_{j=1}^{k_t}\sum_{l=1}^{a_t}\hat{p}_{hj}(t)\hat{p}_{jl}(1)}
                  {\sum_{j=1}^{k_t}\hat{p}_{hj}(t)},
                  & 1\le t \le T-1, \\
         & \frac{\sum_{j=1}^{a_{T}}\hat{p}_{hj}(T)}{\sum_{l=1}^{k_{T}}\hat{p}_{hl}(T)}, &
          t=T,
 \end{aligned}
\right.
\end{eqnarray}
where $a_{t}$ is the largest state index which satisfies
\begin{equation}
 a_{t}=\max
 \left\{
 k \in \left\{ 1,2,\cdots,n  \right\}:  q_k < (P_0-Q_0)(1+r)^{t}
 \right\},
\end{equation}
where $q_k$ is the representative price level for the kth state.
\end{proposition}

\section{Active Margin System}
\subsection{Individualized Maintenance Margin}
% set of indifferent
\citet{Peter2000} showed that there is a great gap between the amount of margin debt outstanding and the maximum allowed margin loans between 1985 and 2000, where margin account owners do not use nearly 60 percent of their debt capacity. This observation indicates that the amount of initial margin deposited by investors is much more than the requirements of authority. On the other hand, the more the initial margin is, the less risk the margin lending transaction is. In order to encourage  investors  to deposit more initial margin, and to compete with other brokers,  the maintenance margin ratios should be relatively smaller for those investors with more initial margin to reduce their requirements of maintenance margin and  frequencies of margin calls.

With the help of CPNR, it is possible  to construct an individualized  margin system for the investor with a specified amount of initial margin. For an individual investor with initial margin $Q_0$, the least maintenance margin ratio $w^*(Q_0)$ which satisfies the constraint of CPNR can be used as the required maintenance margin ratio for this investor, which is called the individualized maintenance margin ratio.
When the amount of initial margin deposited by investor satisfies the condition of Proposition \ref{initialmargin} with respect to the individualized maintenance margin ratio, it is said to be adequate under the constraint of CPNR, otherwise the investor should deposit more initial margin to fulfill the requirement of maintenance margin.

\subsection{Deduced Margin System}
% the squared method
In order to give an example of margin system deduced from the constraint of CPNR, an active
margin system is constructed and applied to Chinese margin lending market in the following sections. The amount of initial margin is specified by the value of initial margin ratio, which  is chosen from 0.01 to 1 with step length 0.01, and the ith ratio is denoted as $m_i$, $i=1,2,\cdots, 100$.
The minimum maintenance margin ratio which satisfies the constraint of CPNR with respect to $Q_i=m_i*P_0$ is denoted as $w_i$,  which is the optimal one that minimizes the opportunity cost of the loan with respect to the same initial margin. The resulted margin system is denoted as a pair of numbers $(m_i,w_i)$, $i=1,2,\cdots, 100$.

For a particular value of CPNR, such as $\mathrm{CPNR}=0.05$, there are several margin systems which satisfy the constraint of CPNR and Proposition \ref{initialmargin} with $1 < w_i \le 1.5$, denoted as $(m_i,w_i)$, $i=1,2,\cdots,q$, where $q$ is the total number of those indifference margin systems.
A specified principle is needed to choose an optimal margin system from the set of indifference margin systems.
In order to give an alternative margin system which can be used to compare with the recommended margin system, a least squares method is adopted  in this paper.
In other words, the optimal margin system $(m^*,w^*)$ should solve the following problem
\begin{eqnarray*}
\min_{(m,w)} \sum_{i=1}^{q} \left( m_i -m \right)^2 +  \left(w_i -w \right)^2, \\
\text{s.t.} (m,w) \in \left\{ (m_j,w_j), j=1,2,\cdots,q  \right\}.
\end{eqnarray*}
The resulted margin system $(m^*,w^*)$ is called the deduced margin system, and the margin system recommended by CSRC and the two exchanges is called the required margin system in the following sections.

\subsection{Dynamics of Margin System}

% insert the example of the dynamic of margin system
\begin{figure}
\caption{The Dynamics of Margin Ratios for SH600123 with CPNR=0.05}\label{dynamic123}
\begin{center}
\includegraphics[width=1.0\textwidth]{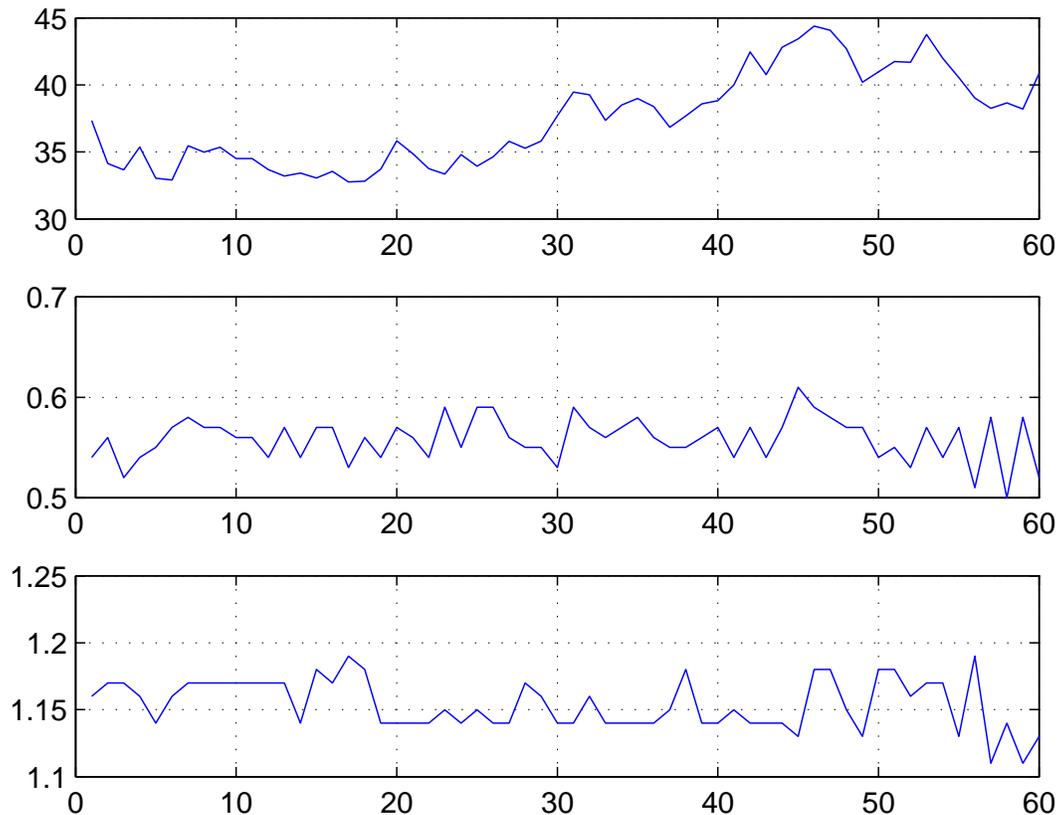}
 \end{center}
 {Notes.
  The top of the figure is the sequence of stock prices of SH600123 between May 8, 2009 and
August 3, 2009, which is listed on the SSE. The middle and the bottom of the figure are the sequences of the deduced initial and  maintenance margin ratios for the margin loans with period of 30 days respectively.}
\end{figure}

In order to explain the dynamics of deduced margin system, an example is given in this subsection.  The stock prices of SH600123 between
May 8, 2009 and
August 3, 2009 are plotted in the top of Figure \ref{dynamic123}, the corresponding initial and maintenance margin ratios are plotted in the middle and  bottom of the figure respectively.
The value of CPNR is chosen to be 0.05 in this example.

From Figure \ref{dynamic123}, it  can be  found that the deduced initial and maintenance margin ratios vary with respect to the changes of stock prices. This observation indicates that
 when the dynamics of market prices is taken into account, the resulted margin system should be  active with respect to the changes of stock prices. And it can be found that the deduced initial margin ratios are likely to increase before the declines in  stock prices. On the other hand, the deduced initial margin ratios are likely to decrease, when there is an upward tendency of stock price movement, such as the situation between the 25th and 30th days. Those observations are consistent with the conclusions about how the margin policy should look like from a theoretical point of view in \citet{ricke_2003}.

\section{Empirical Analysis of Chinese Market}

\subsection{Design of Out-of-Sample Test}
%learing and estimate of ratios
% construction of math model
For the margin loan traded on a specified date, 800 prices  just before the current date  are used to construct the Markov chain and the margin system, where the number of a group is chosen to be 25, and the period of the loan is 30 days in this paper.
30 prices following the current date are used to test whether the loan is protected by the margin system. Whenever a margin call is issued during the period of the loan, collaterals are supposed to be liquidated within one trading day, and the margin system is said to pass the out-of-sample test when the return of the broker is nonnegative after mandatory liquidation. In our empirical investigations, 200 margin loans are constructed for each stock, and a stock is said to pass the out-of-sample test when its frequency of negative returns among those margin loans is less than or equal to the specified value of CPNR.

% range of sample size
As the sample size needed in  this paper is at least 1030, the stocks used in the empirical investigations should be listed before 2007.  There are 134 stocks in the SSE 180 Index which have been trading since 2007,  among which 119 stocks pass the out-of-sample test in our empirical investigations with CPNR=0.05. There are more 8 stocks which can pass the test when the sample size used to construct the Markov chain is increased from 800 to 1,500. In order to investigate the performance of deduced margin system with the same depth of memory, the empirical investigations in the following subsections  are limited to those 119 stocks which have passed the out-of-sample test, using 800 historical data to construct the Markov chain model.

\subsection{Initial Margin Ratio}
% there is the table of initial margin ratio
% table of initial margins
\begin{table}
\caption{Quantile Analysis of the Initial Margin Ratios  under the Deduced Margin System} \label{initialtable}
\begin{center}
\begin{tabular*}{15cm}{@{\extracolsep{\fill}}clllllll}
\hline
        \textit{Statistics with} &   \multicolumn{3}{c}{ } & \multicolumn{4}{c}{Quantiles}  \\
        \cline{5-8}
\textit{119 Observations}    &   min &   max &   mean    &   0.70    &   0.80    &   0.90    &   0.95     \\ \hline
\textit{minimum}    &   0.27    &   0.57    &   0.49    &   0.50    &   0.51     &  0.52     &  0.55    \\
\textit{maximum }   &   0.59    &   0.85    &   0.66    &   0.66    &   0.67     &  0.71     &  0.76    \\
\textit{mean }  &   0.53    &   0.61    &   0.57    &   0.58    &   0.58    &    0.59    &  0.60    \\
\textit{quantiles}  &       &       &       &       &       &       &       \\
0.20    &   0.45    &   0.59    &   0.54    &   0.55    &   0.55    &   0.57     &  0.58    \\
0.30    &   0.46    &   0.60    &   0.55    &   0.56    &   0.57    &   0.58     &  0.59    \\
0.40    &   0.46    &   0.61    &   0.56    &   0.57    &   0.58    &   0.59     &  0.60    \\
0.50    &   0.49    &   0.61    &   0.57    &   0.57    &   0.59    &   0.60     &  0.61    \\
0.60    &   0.54    &   0.63    &   0.58    &   0.58    &   0.59    &   0.60     &  0.61    \\
0.70    &   0.55    &   0.64    &   0.58    &   0.59    &   0.61    &   0.61     &  0.62    \\
0.80    &   0.56    &   0.66    &   0.60    &   0.61    &   0.62    &   0.62     &  0.63    \\
0.90    &   0.56    &   0.79    &   0.61    &   0.62    &   0.63    &   0.64     &  0.66    \\
0.95    &   0.57    &   0.80    &   0.62    &   0.63    &   0.64    &   0.66     &  0.70    \\
\hline
\end{tabular*}
\end{center}
\vskip 0.8cm
 {Notes. 12 statistics of initial margin ratios are observed for each stock, including the minimum value, maximum value, mean, and 20\%, 30\%, 40\%, 50\%, 60\%, 70\%, 80\%, 90\%, 95\% quantiles respectively. And the minima, maxima, mean, and  70\%, 80\%, 90\%, 95\% quantiles are reported for each statistic respectively.
}
\end{table}

As 200 margin loans are constructed for each stock, there are 200 observations of initial margin ratios for every stock under consideration. In order to investigate the distributional properties of those initial margin ratios, a quantile analysis is applied to the
observations. 12 statistics are observed from the initial margin ratios of each stock, including the minimum value, maximum value, mean, and 20\%,  30\%, 40\%, 50\%, 60\%, 70\%, 80\%, 90\%, 95\% quantiles respectively, therefore there are 119 observed values for each statistic. A quantile analysis is also applied to the observations of each statistic, including the minima, maxima, mean and 70\%, 80\%, 90\%, 95\% quantiles  respectively. The results are reported in Table $\ref{initialtable}$.

From Table $\ref{initialtable}$, we can find that the minima of the statistic {\it minimum} is $27\%$, and its  maxima and  mean are $57\%$, $49\%$ respectively. The $70\%$ quantile of those observed minimum initial margin ratios is $50\%$, which means that there are only $30\%$ stocks considered in this paper whose minimum initial margin ratios are larger than or equal to $50\%$, which is the recommended minimum value by the stock exchanges and CSRC. The mean values of those 9 observed quantiles  are distributed between $54\%$ and $62\%$, which indicates that  the initial margin ratios for most of the stocks considered in this paper are larger than $50\%$ under the deduced margin system.

\subsection{Maintenance Margin Ratio}
% 130 as the limitation is too large for many stock, should be differently for
% each stock

% table of maintenance margins
\begin{table}
\caption{Quantile Analysis of the Maintenance Margin Ratios under the Deduced Margin System}
\begin{center}
\begin{tabular*}{15cm}{@{\extracolsep{\fill}}clllllll}
\hline
    \textit{Statistics with} &   \multicolumn{3}{c}{ } & \multicolumn{4}{c}{Quantiles}  \\
    \cline{5-8}
\textit{119 Observations}     &   min &   max &   mean    &   0.70    &   0.80    &   0.90    &   0.95     \\ \hline
\textit{minimum}    &   1.00    &   1.25    &   1.09    &   1.13    &   1.15     &  1.17    &   1.19    \\
\textit{maximum }   &   1.17    &   1.69    &   1.33    &   1.38    &   1.43     &  1.46    &   1.49    \\
\textit{mean }  &   1.11    &   1.42    &   1.21    &   1.23    &   1.26    &    1.30   &   1.34    \\
\textit{quantiles}  &       &       &       &       &       &       &       \\
0.20    &   1.00    &   1.36    &   1.17    &   1.19    &   1.20    &   1.24     &  1.28    \\
0.30    &   1.00    &   1.38    &   1.18    &   1.20    &   1.22    &   1.26     &  1.30    \\
0.40    &   1.11    &   1.40    &   1.20    &   1.21    &   1.24    &   1.28     &  1.32    \\
0.50    &   1.11    &   1.42    &   1.21    &   1.23    &   1.26    &   1.30     &  1.35    \\
0.60    &   1.12    &   1.44    &   1.23    &   1.24    &   1.28    &   1.32     &  1.37    \\
0.70    &   1.12    &   1.45    &   1.24    &   1.26    &   1.29    &   1.34     &  1.39    \\
0.80    &   1.12    &   1.47    &   1.26    &   1.28    &   1.31    &   1.37     &  1.41    \\
0.90    &   1.13    &   1.50    &   1.28    &   1.31    &   1.35    &   1.40     &  1.44    \\
0.95    &   1.14    &   1.64    &   1.30    &   1.33    &   1.38    &   1.42     &  1.46    \\

\hline
\end{tabular*}
\end{center}
\vskip 0.8cm
{Notes.
12 statistics of maintenance margin ratios are observed for each stock, including the minimum value, maximum value, mean, and 20\%, 30\%, 40\%, 50\%, 60\%, 70\%, 80\%, 90\%, 95\% quantiles respectively. And the minima, maxima, mean, and  70\%, 80\%, 90\%, 95\% quantiles are reported for each statistic respectively.
}\label{maintmargin}
\end{table}

A similar quantile analysis is applied to those observed statistics for maintenance margin ratios. The minimum value, maximum value, mean and $9$ quantiles are observed for each stock respectively, and the distributional properties of each statistic are investigated through their minima, maxima, mean, and   $70\%$, $80\%$, $90\%$, $95\%$ quantiles respectively. The results are reported in Table \ref{maintmargin}.

From Table  \ref{maintmargin}, we can find that the average of those observed means is $121\%$, which is smaller than the recommended value $130\%$ given by the two Chinese stock exchanges and CSRC. The $90\%$ quantile of those observed  $50\%$ quantiles is $130\%$, which indicates that there are only $10\%$ stocks considered in this paper whose $50\%$ quantiles of maintenance margin ratios are larger than $130\%$. And the $80\%$ quantile of those observed  $80\%$ quantiles is $131\%$, which indicates that the maintenance margin ratios for most of the stocks considered in this paper are  below $130\%$ under the deduced margin system.

\subsection{Numbers of  Margin Calls}
% analysis
% table of initial margins
\begin{table}
\caption{Quantile Analysis of the Numbers of Margin Calls among 200 Margin Loans for Each Stock}
\begin{center}
\begin{tabular*}{15cm}
{@{\extracolsep{\fill}}rrrrrrrrrrrr}
\hline
 &   \multicolumn{3}{c}{ } & \multicolumn{6}{c}{Quantiles} \\
 \cline{5-10}
  &   min &   max &   mean    &   0.30    &   0.50    &   0.80    &   0.90     &  0.95    &   0.99    \\ \hline
\textit{Required Margin}    &   0.00    &   70.00   &   22.30   &   10.40   &    23.00  &   34.00   &   45.20   &   54.10   &   67.93   \\
\textit{Deduced Margin} &   0.00    &   10.00   &   0.97    &   0.00    &   0.00     &  2.00    &   4.00    &   6.00    &   8.62    \\
\hline
\end{tabular*}
\end{center}
\vskip 0.8cm
{Notes.
The numbers of margin calls among 200 margin loans of each stock are reported in this table under the two types of margin systems. The results under the required margin system are reported in the line of \textit{Required Margin}, and the results given by the deduced margin system are reported in the line of \textit{Deduced Margin}.
}\label{callnumber}
\end{table}

% table of initial margins
\begin{table}
\caption{Quantile Analysis of the Costs for Margin Loans under the Two Types of Margin Systems}
\begin{center}
\begin{tabular*}{15cm}{@{\extracolsep{\fill}}c rrrrrrrc}
\hline
\textit{Statistics with} &   \multicolumn{3}{c}{ } & \multicolumn{4}{c}{Quantiles} & \\
\cline{5-8}
 \textit{119 Observations}    &   min &   max &   mean    &   0.70    &   0.80    &   0.90    &   0.95     &  RD  \\
\hline
\textit{minimum}    &   1.72    &   60.91   &   8.04    &   8.63    &   10.39    &  13.18   &   19.03   &       \\
    &   1.80    &   56.55   &   7.53    &   8.19    &   9.35    &   12.28   &    18.13  &   0.05    \\
\textit{maximum}    &   3.62    &   103.84  &   15.30   &   16.96   &   19.94    &  25.68   &   30.96   &       \\
    &   2.87    &   88.74   &   14.84   &   16.21   &   19.19   &   25.15   &    30.94  &   0.00    \\
\textit{mean}   &   2.55    &   86.87   &   11.08   &   12.40   &   14.50   &    19.47  &   23.82   &       \\
    &   2.19    &   79.37   &   9.94    &   11.10   &   12.59   &   17.34   &    21.78  &   0.09    \\
\textit{quantiles}  &       &       &       &       &       &       &       &        \\
0.20    &   2.35    &   81.70   &   9.80    &   10.94   &   12.74   &    17.86   &  21.60   &       \\
    &   2.02    &   74.24   &   8.77    &   9.39    &   11.11   &   15.33   &    20.08  &   0.08    \\
0.30    &   2.40    &   84.03   &   10.24   &   11.32   &   13.28   &    18.46   &  22.36   &       \\
    &   2.05    &   78.01   &   9.15    &   9.77    &   11.53   &   15.76   &    20.84  &   0.07    \\
0.40    &   2.44    &   86.00   &   10.63   &   11.74   &   13.87   &    19.00   &  23.26   &       \\
    &   2.10    &   80.33   &   9.48    &   10.20   &   11.93   &   16.50   &    21.47  &   0.08    \\
0.50    &   2.49    &   88.07   &   11.01   &   12.16   &   14.34   &    19.54   &  23.69   &       \\
    &   2.14    &   81.76   &   9.80    &   10.60   &   12.55   &   16.95   &    21.83  &   0.08    \\
0.60    &   2.58    &   88.95   &   11.37   &   12.59   &   14.76   &    20.02   &  24.34   &       \\
    &   2.21    &   82.73   &   10.15   &   11.17   &   12.96   &   17.47   &    22.23  &   0.09    \\
0.70    &   2.67    &   92.04   &   11.79   &   13.35   &   15.16   &    20.77   &  24.96   &       \\
    &   2.24    &   83.49   &   10.50   &   11.75   &   13.53   &   18.36   &    22.63  &   0.10    \\
0.80    &   2.76    &   94.11   &   12.34   &   13.80   &   15.80   &    21.47   &  25.79   &       \\
    &   2.31    &   84.66   &   10.93   &   12.34   &   14.12   &   19.13   &    23.27  &   0.11    \\
0.90    &   2.86    &   96.27   &   13.07   &   14.60   &   17.09   &    22.20   &  26.57   &       \\
    &   2.46    &   85.74   &   11.58   &   12.97   &   14.77   &   20.54   &    23.97  &   0.11    \\
0.95    &   3.01    &   97.68   &   13.65   &   15.09   &   18.19   &    22.83   &  27.34   &       \\
    &   2.63    &   87.07   &   12.46   &   13.96   &   15.70   &   21.84   &    25.58  &   0.07    \\
\hline
\end{tabular*}
\end{center}
\vskip 8pt
{Notes.
The first line of each statistic are observations under the deduced margin system, and the second line are observations under the required margin system respectively. The last column of this table are the relative differences (RD) between  those two 95\% quantiles with respect to the corresponding values under the required margin system  respectively.
}\label{loancost}
\end{table}

Mandatory liquidation is the last defence line against the risk associated with transactions of margin lending,
 however, frequent margin calls may destabilize stock prices,
   therefore the numbers of margin calls under the two types of margin systems  are investigated in this subsection.
   There are 200 margin loans for each stock considered in this paper, and the numbers of loans which have to meet a margin call are calculated for each stock under the required and the deduced margin systems respectively. A quantile analysis is applied to the observed numbers, and the results are reported in Table \ref{callnumber}.

The observations under the required margin system are reported in the line of \textit{Required Margin}, and the results under the deduced margin system are reported in the line of  \textit{Deduced Margin} respectively. From Table \ref{callnumber}, it can be found that the mean of the 119 observed numbers of margin calls is 22.30 under the required  margin system, which indicates that there are averagely 22.3 margin calls among  200 margin loans for each stock. And its counterpart under the deduced margin system is 0.97, which indicates that there are averagely only 0.97 margin calls among 200 margin loans for each stock. It can be concluded that the number of margin calls under the deduced margin system is much less than its counterpart under the required  margin system.

It is notable that the frequencies of negative returns under these two types of margin systems are both less than or equal to 5\% for those stocks considered in this subsection, therefore the reduction in the frequency of margin calls under the deduced margin system is under an acceptable level of risk, which may be beneficial to the stability of stock market.

\subsection{Cost of Margin Loan}
% analysis

The cost of a margin loan is the terminal time value of the initial margin and the additional capital deposited to broker to meet all of the margin calls during the period of the loan. The cost of a margin loan is determined by the amount of initial margin, market value of the loan, required maintenance margin ratio, riskless interest rate, and the evolution of stock prices during the period of the loan.
There are 200 margin loans for each stock considered in this paper, and a similar quantile analysis is applied to the observed costs under the two types of margin systems respectively.

Table \ref{loancost} reports the results of quantile analysis. The elements in the first row of each statistic are the observations under the
deduced  margin system, and the second row are those observations given by the
required  margin system. The last value of each block is the relative  differences (RD) between those two $95\%$ quantiles with respect to the corresponding values under the required  margin system. Those observed RDs indicate that
the average cost of each margin loan under the deduced margin system
is about $10\%$ larger than its counterpart under the required margin system. Therefore it can be concluded  that the frequency of margin calls is  reduced by at least $90\%$ with more $10\%$ cost under the deduced margin system.

\citet{Peter2000} showed that there is a great gap between the amount of margin debt outstanding and the amount  investors are maximally able to  borrow between 1985 and 2000, where the margin account owners do not use nearly 60 percent of their debt capacity. This observation indicates that the amount of initial margin that investors deposit to their brokers are much more than the required amount of the authorities, therefore
 the more $10\%$ cost under the deduced margin system than its counterpart under the required margin system is acceptable in practice.

\section{Discussions and Conclusions}

The margin system for margin lending transactions  in China is composed of initial margin requirement, maintenance margin requirement and mandatory liquidation, which is the line of defence against the risk associated with the transactions of margin lending.
In order to react to the changes of stock prices, and to protect  brokers from customer defaults with a clearly specified target of risk control,
 an active  margin system is proposed in this paper, where the conditional probability of negative return (CPNR) is used to measure the risk faced by brokers.
The resulted margin system can be individualized for the investors with different amounts of initial margin.

An example of active margin system is given in the previous sections, which is chosen from an indifference set of margin systems with the same value of CPNR by a least squares method. The resulted margin system is applied to the margin loans of 134 stocks listed on the SSE. There are 119 stocks which pass the out-of-sample test, where 200 margin loans with a 30-day period are constructed for each stock. In order to investigate the distributional properties of  initial and maintenance margin ratios, numbers of margin calls, and the costs of margin loans, a quantile analysis is applied to those observations from the stocks which have  passed  the out-of-sample test.

It is found that the initial margin ratios under the deduced margin system are generally larger than $50\%$, which is consistent with the observations that the initial margins deposited by investors are generally larger than the recommended values of the authorities. And the maintenance margin ratios are generally less than $130\%$, which is the value of minimum maintenance margin ratio recommended by CSRC and both of Chinese stock exchanges. The average cost of margin loans under the deduced margin system is  $11.1\%$ larger than its counterpart under the required margin system, while the corresponding frequencies of margin calls are reduced $95.7\%$ under the deduced margin system  with respect to  its counterpart under the required margin system. Those empirical investigations indicate that the idea of CPNR can be used to set an operational margin system for margin lending transactions.

% markov model should be more complex
CPNR is an idea of risk measurement focusing on the market value of collaterals and the  stop loss function of mandatory liquidation,
which may be realized under several candidates of working models, such as the geometric Brownian motion, fractional geometric Brownian motion,  and the GARCH models etc.. And the method  used to reconstruct the Markov version of stock prices  is relatively simple in this paper. Those topics about the working models will be left for further research.

\begin{APPENDIX}{Proof of the Main Results}
\proof{Proof of Proposition 3}
Notice that~$B_i\cap B_j=\O$, if $i\neq j$, and $B=\sum_{i=1}^{T}B_i$,  we have
\begin{equation}
\mathrm{Prob}(B)=\sum_{i=1}^T \mathrm{Prob}(B_i).
\end{equation}
And it is directly from the definition of $B_i$ and the  Markov property of stock price, we have the following probability
\begin{equation}
\mathrm{Prob}(B_t)=\mathrm{Prob}(\overline{D}_1)
\mathrm{Prob}(\overline{D}_2|\overline{D}_1)\cdots{\mathrm{Prob}(\overline{D}_{t-1}|
\overline{D}_{t-2})}\mathrm{Prob}({D_{t}}|\overline{D}_{t-1}).
\end{equation}
Let $k_m$ denote the largest state index which satisfies
\begin{equation}
 k_m = \max \left\{ k \in \left\{ 1,2,\cdots,n \right\} : q_{k} < (wP_0-Q_0)(1+r)^{m} \right\},
\end{equation}
where $q_k$ is the representative price level for the kth state. Let $s_h$ denote   the state of the current stock price,
\begin{eqnarray}
 &\mathrm{Prob}(B_1)&=\mathrm{Prob}(D_1)\nonumber \\
 & & =P\{P_1\in \left\{ s_1, s_2, \cdots, s_{k_1} \right\}|P_0\in s_h\}\nonumber \\
 & &=\sum_{i=1}^{k_1}p_{hi}(1).
\end{eqnarray}
From the Markov property of stock price, we have
\begin{eqnarray}\
 \mathrm{Prob}\left(D_m|\overline{D}_{m-1}\right)
 & = &
  \frac{\mathrm{Prob}\left( \overline{D}_{m-1}D_m\right)}
   {\mathrm{Prob}\left(\overline{D}_{m-1}\right)}\nonumber \\
 & = & \frac{\mathrm{Prob}\left\{ P_m \in \left\{s_1,\cdots,s_{k_m}\right\}, P_{m-1}\in \left\{s_{k_{m-1}+1},\cdots,s_{n}\right\}|P_0 \in s_h\right\}}
 {\mathrm{Prob}\left\{P_{m-1}\in \left\{
 s_{k_{m-1}+1},s_{k_{m-1}+2},\cdots,s_n\right\}|P_0 \in s_h\right\}}\nonumber \\
 & = & \frac{\sum_{i=k_{m-1}+1}^n \sum_{j=1}^{k_m} p_{hi}(m-1)p_{ij}(1)}{\sum_{i=k_{m-1}+1}^{n}p_{hi}(m-1)},
\end{eqnarray}
which finishes the proof of Proposition 3.\Halmos
\endproof

\proof{Proof of Proposition 4}
From the definition of $A$, $B$, and $B_i$, we have the following formula
\begin{equation}
 \mathrm{Prob} \left( AB \right) = \sum_{i=1}^T \mathrm{Prob}\left( AB_i\right).
\end{equation}
And from the definition of $B_i$, $D_i$ and the Markov property of stock price, we have
\begin{eqnarray}\
 \mathrm{Prob}(AB_i)&=&\mathrm{Prob}\left( A\overline{D}_1 \overline{D}_2\cdots\overline{D}_{i-1}D_i \right)\nonumber \\
  &=&\mathrm{Prob}\left(\overline{D}_1\right)
  \mathrm{Prob}\left( \overline{D}_2|\overline{D}_1\right)
  \cdots{\mathrm{Prob}\left( \overline{D}_{i-1}|\overline{D}_{i-2}\right)}\nonumber \\
  & & \mathrm{Prob}\left( {D_{i}}|\overline{D}_{i-1}\right)\mathrm{Prob}\left( A|D_{i}\right).
\end{eqnarray}
If $1 \le i \le T-1$, we have
\begin{eqnarray}
 &\mathrm{Prob}\left( A|D_{i}\right)
  & = \frac{\sum_{j=1}^{k_i}\sum_{l=1}^{a_{i}}p_{hj}(i)p_{jl}(1)}
   {\sum_{j=1}^{k_i}p_{hj}(i)},
\end{eqnarray}
where $a_i$ is the state index which satisfies
\begin{equation}
a_i = \max \left\{ k \in
       \left\{1,2,\cdots,n  \right\}: q_k < \left( P_0 - Q_0 \right)(1+r)^i
          \right\}.
\end{equation}
Otherwise $i=T$, we have
\begin{eqnarray}
 &\mathrm{Prob}\left( A|D_{i}\right)
  & = \frac{\sum_{j=1}^{a_T}p_{hj}(T)}
   {\sum_{j=1}^{k_T}p_{hj}(T)},
\end{eqnarray}
which finishes the proof of Proposition 4.\Halmos
\endproof
\end{APPENDIX}

% Appendix here
% Options are (1) APPENDIX (with or without general title) or
%             (2) APPENDICES (if it has more than one unrelated sections)
% Outcomment the appropriate case if necessary
%
% \begin{APPENDIX}{<Title of the Appendix>}
% \end{APPENDIX}
%
%   or
%
% \begin{APPENDICES}
% \section{<Title of Section A>}
% \section{<Title of Section B>}
% etc
% \end{APPENDICES}

% Acknowledgments here
\ACKNOWLEDGMENT{Guanghui Huang is supported by the Fundamental Research Funds for the Central Universities of China, project No. CDJZR10 100 007.}

% References here (outcomment the appropriate case)

% CASE 1: BiBTeX used to constantly update the references
%   (while the paper is being written).
%\bibliographystyle{ormsv080} % outcomment this and next line in Case 1
%\bibliography{<your bib file(s)>} % if more than one, comma separated

% CASE 2: BiBTeX used to generate mypaper.bbl (to be further fine tuned)
%\input{mypaper.bbl} % outcomment this line in Case 2

%If you don't use BiBTex, you can manually itemize references as shown below.

\bibliographystyle{nonumber}

%%%%%%%%%%%%%%%%%
\end{document}